# A catalogue of faint local radio AGN and the properties of their host galaxies

E. K. Lofthouse[1]*, S. Kaviraj[1], D. J. B. Smith[1] and M. J. Hardcastle[1]

[1]*Centre for Astrophysics Research, School of Physics, Astronomy and Mathematics, University of Hertfordshire, College Lane, Hatfield, AL10 9AB, UK*

21 May 2018

**ABSTRACT**

We present a catalogue of local ($z < 0.1$) galaxies that contain faint AGN. We select these objects by identifying galaxies that exhibit a significant excess in their radio luminosities, compared to what is expected from the observed levels of star-formation activity in these systems. This is achieved by comparing the optical (spectroscopic) star formation rate (SFR) to the 1.4 GHz luminosity measured from the FIRST survey. The majority of the AGN identified in this study are fainter than those in previous work, such as in the Best and Heckman (2012) catalogue. We show that these faint AGN make a non-negligible contribution to the radio luminosity function at low luminosities (below $10^{22.5}$ W Hz$^{-1}$), and host ~13 per cent of the local radio luminosity budget. Their host galaxies are predominantly high stellar-mass systems (with a median stellar mass of $10^{11} M_\odot$), are found across a range of environments (but typically in denser environments than star-forming galaxies) and have early-type morphologies. This study demonstrates a general technique to identify AGN in galaxy populations where reliable optical SFRs can be extracted using spectro-photometry and where radio data are also available so that a radio excess can be measured. Our results also demonstrate that it is unsafe to infer SFRs from radio emission *alone*, even if bright AGN have been excluded from a sample, since there is a significant population of faint radio AGN which may contaminate the radio-derived SFRs.

## 1 INTRODUCTION

Active galactic nuclei (AGN) are the result of supermassive black holes at the centres of galaxies accreting material and emitting large amounts of radiation (Lynden-Bell 1969). In the local Universe, there is strong evidence that the evolution of black holes and their host galaxies may be linked, despite the fact that supermassive black holes account for a vanishingly small fraction of the galaxy's mass (e.g. Magorrian et al. 1998). For example, a tight positive correlation exists between black-hole mass and host-galaxy bulge mass (Marconi & Hunt 2003; Häring & Rix 2004) or velocity dispersion (Ferrarese & Merritt 2000; Gebhardt et al. 2000). In addition, the growth of black-holes and the growth of stellar mass show remarkable similarities across cosmic time (Boyle & Terlevich 1998), with the number density of AGN peaking around $1 < z < 3$ (Hasinger et al. 2005; Aird et al. 2010; Delvecchio et al. 2014), which is comparable to the peak in the cosmic star-formation rate density of the Universe (Madau et al. 1996; Lilly et al. 1996; Hopkins & Beacom 2006; Sobral et al. 2013; Madau & Dickinson 2014). Such correlations support the hypothesis that galaxies and their central black holes may grow in lockstep with each other (e.g. Richstone et al. 1998; Haehnelt et al. 1998; Monaco et al. 2000; Kaviraj et al. 2012). This may be the result of a common mechanism driving the growth of both the galaxy and the black hole (Vito et al. 2014) and/or the result of feedback from AGN that regulates the star formation process in the host galaxy (e.g. Silk & Rees 1998).

Feedback from AGN is frequently included in models of galaxy evolution, in order to successfully replicate the observed properties of galaxies e.g. the high-luminosity end of the galaxy luminosity function and rest-frame colours (Di Matteo et al. 2005; Croton et al. 2006; Cattaneo et al. 2009; Sijacki et al. 2015; Dubois et al. 2016; Kaviraj et al. 2017). From a theoretical perspective, AGN are an attractive source of energetic feedback, because the energy released from the process of accretion can be several orders of magnitude larger than the binding energy of the gas reservoir (e.g. Kaviraj et al. 2011), even in the most massive galaxies (Fabian 2012). However, the details of this process remain uncertain and the observational picture is unclear. While some observational studies have offered evidence for AGN reducing star formation activity (Fabian 2012; Tombesi et al. 2015), others have found opposite results, i.e. that AGN either do not appear to strongly regulate star formation activity or that higher luminosity AGN actually appear to exhibit higher star formation rates (Mullaney et al. 2012a,b; Rosario et al. 2013; Smethurst et al. 2015; Gürkan et al. 2015; Kaviraj et al. 2015; Sarzi et al. 2016; Nedelchev et al. 2017).

* E-mail: e.k.lofthouse@herts.ac.uk

To fully understand the evolution of galaxies, and the importance of AGN in shaping their evolution, it is necessary to have complete samples of AGN across cosmic time. At low redshift, the Best & Heckman (2012, hereafter BH12) study has provided a benchmark catalogue of AGN. The BH12 catalogue was created by cross-matching the SDSS DR7 with the Faint Images of the Radio Sky at Twenty centimeters (FIRST) survey and the NRAO (National Radio Astronomy Observatory) VLA (Very Large Array) Sky Survey (NVSS). By combining three AGN selection techniques - optical emission-line diagnostic diagrams (BPT diagram), a comparison of radio and H$\alpha$ luminosities, and the 4000Å break, Best & Heckman (2012) catalogued 1245 radio-loud AGN at $z < 0.1$. While BH12 provide large numbers of AGN, their flux limit of 5 mJy means that there are (possibly many more) fainter AGN that do not appear in this catalogue. For example, using LOFAR data, Gürkan et al. (submitted) have found a sample of sources that have enhanced radio emission compared to what would be expected from star formation alone and yet have not been previously identified as AGN. The authors find that the radio spectra of these objects can be explained by compact AGN and conclude that they are previously uncatalogued low-luminosity radio-loud AGN. This is consistent with the work of Hodge et al. (2008) who median-stacked radio cutouts for optically unclassifiable SDSS galaxies and found conflicting SFRs, indicating that the stacked sample was contaminated by AGN. Various other works are using new techniques to detect previously unidentified AGN. For example, Truebenbach & Darling (2017) use a mid-infrared and radio selection method to identify a catalogue of 46,000 optically-faint AGN.

In this work, we construct a catalogue of galaxies which includes fainter AGN that do not appear in the BH12 catalogue and which, as we show in Section 4, contribute measurably to the local radio luminosity budget at low luminosities. This paper is structured as follows. In Section 2 we describe the data used in this work. In Section 3 we describe the construction of our AGN catalogue. In Section 4 we study the contribution of this sample to the local radio luminosity function and in Section 5 we investigate the properties of the host galaxies of these AGN e.g. their stellar mass (Section 5.1), local environment (Section 5.2) and morphologies (Section 5.3). We summarize our findings in Section 6. Throughout we use the following cosmological parameters $\Omega_m = 0.309$, $\Omega_\Lambda = 0.691$ and $H_0 = 68$ kms$^{-1}$ Mpc$^{-1}$ (Planck Collaboration et al. 2016).

## 2 DATA

We select our galaxy sample by cross-matching the public seventh data release (DR7; Abazajian et al 2009) of the Sloan Digital Sky Survey (SDSS) with the Faint Images of the Radio Sky at Twenty-cm survey (FIRST survey; Becker et al. 1995), restricted to the local Universe ($z < 0.1$). The DR7 release is chosen so that the results are consistent with other the catalogues used in this work, as described below. The FIRST survey maps over 10,000 square degrees of the sky, chosen to coincide with the SDSS, using the NRAO Very Large Array at 1.4 GHz. The optical and radio sources are matched using a 4 arcsecond radius which is chosen to maximise matches while being small enough to avoid large numbers of false positives. While the FIRST resolution is 5 arcseconds, a 4 arcsecond radius still selects 98% of the matches found using a 5 arcsecond radius. Due to the similarity in the results between these radii, we choose to use the slightly more conservative 4 arcseconds. This optical-radio matching results in a sample of 12,052 galaxies at $z < 0.1$. Given that there are 381,798 SDSS sources at $z < 0.1$ and the SDSS footprint is 15,000 square degrees, we calculate the density of sources to be 25.4 per square degree. Therefore we expect less than 340 false matches within our full sample.

Our methods for obtaining SFR, stellar mass, environmental and morphological data closely follow that of Lofthouse et al. (2017). We provide a summary here but refer readers to this paper for further details. Aperture-corrected star formation rates (SFRs) and stellar masses for each galaxy are obtained from the MPA-JHU catalogue[1], which are calculated using SDSS spectra and photometry (Kauffmann et al. 2003; Brinchmann et al. 2004, B04 hereafter). B04 use a methodology based on Charlot et al. (2002). This employs the models from Charlot & Longhetti (2001), which incorporate the Bruzual & Charlot (1993) stellar models, along with emission-line modelling from CLOUDY (Ferland 1996). B04 create model grids with various free parameters, including metallicity and total V-band dust attenuation, and use a Bayesian approach to find the likelihood of each model. Dust attenuation is accounted for in the SFR estimates, following the method of Charlot & Fall (2000), which provides a multi-component dust model that incorporates stellar birth clouds with a finite lifetime (and does not rely on the simplified assumption of dust originating from a foreground screen). B04 account for possible AGN contamination by using different SFR diagnostics depending on the position of the galaxy in the BPT diagram. For the composite, AGN and unclassified galaxies, the SFR is calculated from the 4000Å break. Note that this method applies a dust correction, similar to the average for a star-forming galaxy. However, where the galaxies have S/N > 3 in both H$\alpha$ and H$\beta$, the probability distribution functions are constructed from galaxies with similar H$\alpha$/H$\beta$ to remove all trends with dust attenuation. We direct readers to Salim et al. (2007) and B04 for more details of the procedure.

As described in Lofthouse et al. (2017), we use the environment catalogue of Yang et al. (2007), to determine the local environment of each galaxy in our sample, where galaxies in the 'Field' are classified as those with dark-matter halo masses less than $10^{13}$M$_\odot$, galaxies in 'Groups' have halo masses between $10^{13}$M$_\odot$ and $10^{14}$M$_\odot$ and 'Cluster' galaxies have halo masses above $10^{14}$M$_\odot$. Cross-matching to the Yang et al. catalogue yields 11,313 galaxies in the $z < 0.1$ galaxy sample each of which have measured optical and radio luminosities, masses and local environments.

For our galaxy morphologies, we use the catalogue of bulge-disc decompositions from Lackner & Gunn (2012). Cross-matching this catalogue to the SDSS-FIRST $z < 0.1$ galaxy sample yields 1971 galaxies with optical and radio luminosities, stellar masses and estimates of morphology. This restricted sample (due to the lower redshift limit in the Lackner & Gunn (2012) catalogue of $z < 0.05$) is only used when studying the morphology of the galaxy hosts in Section 5.3.

---

[1] http://www.mpa-garching.mpg.de/SDSS/DR7/

## 3 AGN SELECTION

We select our AGN by identifying galaxies that exhibit an excess in their radio luminosities compared to what might be expected from the observed level of star formation activity in these systems. Fig. 1 shows a comparison of the optical SFRs from the MPA-JHU catalogue (derived using SDSS observations as described above), with the FIRST 1.4 GHz luminosities. In galaxies where the radio luminosity is driven purely by star formation, the optical SFRs and radio luminosity will be strongly correlated (blue dotted line). For comparison, we also show the 'radio-equivalent SFR' on the top axis, calculated from the measured FIRST 1.4 GHz luminosity. These SFRs (and hence the radio-SFR enhancement required to be included in our selection) depend on the choice of radio-SFR calibration. In this work we employ the linear relation of Condon (1992):

$$\text{SFR}_{\text{RADIO\_EQV}} M_\odot \text{yr}^{-1} = \frac{L_{1.4\text{GHz}}}{4 \times 10^{21} \text{W Hz}^{-1}} \quad (1)$$

Note that some papers (e.g. Schmitt et al. 2006; Murphy et al. 2011) have found that this calibration may overestimate the SFRs by around a factor of 2. For the analysis in this paper we use the observed 1.4 GHz luminosity and therefore the uncertainty in the radio-SFR calibrations will not affect our results.

A large fraction of galaxies fall around the line of equivalence where the radio luminosity is consistent with star formation (i.e. the optical SFR is in agreement with the radio-equivalent SFR). However, there is a second population of galaxies which are clearly offset from this line, i.e. where we observe a significant radio luminosity excess beyond what is expected based on their spectroscopic SFRs. This radio excess is a result of a contribution to the radio luminosity by the AGN (e.g. Shabala et al. 2012). Since the optical SFRs are dust corrected, the discrepancy is not driven by dust suppression of optical emission.

We define our AGN as galaxies that fall below the green dotted line which separates the purely star-forming galaxies from the radio excess galaxies. The cut is defined as

$$\log_{10} L_{1.4\text{GHz}} > \log_{10} \text{SFR}_{\text{OPTICAL}} + 22.8 \quad (2)$$

which equates to a radio equivalent SFR of

$$\log_{10} \text{SFR}_{\text{RADIO\_EQV}} > \log_{10} \text{SFR}_{\text{OPTICAL}} + 1.2 \quad (3)$$

This cut results in a sample of 3,510 galaxies potentially hosting AGN, with 8,542 galaxies to the left of the green dotted line which we classify as star-forming (i.e. not hosting AGN). Galaxies close to this boundary may still contain small contributions from very faint AGN. In particular, of the 8,542 galaxies we classify as star-forming, 37 are identified by BH12 as AGN. As shown in Fig. 1, where the red contour of the BH12 AGN overlaps the selection cut, these galaxies lie close to the boundary. Therefore they likely contain some very faint AGN component but their radio-enhancement relative to the optical SFRs is not significant enough to be selected by our method.

37 per cent of the 3,510 AGN thus selected are already included in the BH12 catalogue. This leaves 2,210 galaxies (63 per cent of the galaxies selected by the cut) which we identify as AGN but which are not included in BH12. These galaxies still exhibit enhanced radio emission, indicating the presence of an AGN, but typically have lower luminosities than the BH12 sample. Hereafter, we refer to this sample of 2,210 galaxies as 'faint AGN'. We release this sample of faint AGN with this paper and show the first 20 objects in this catalogue in Table 1.

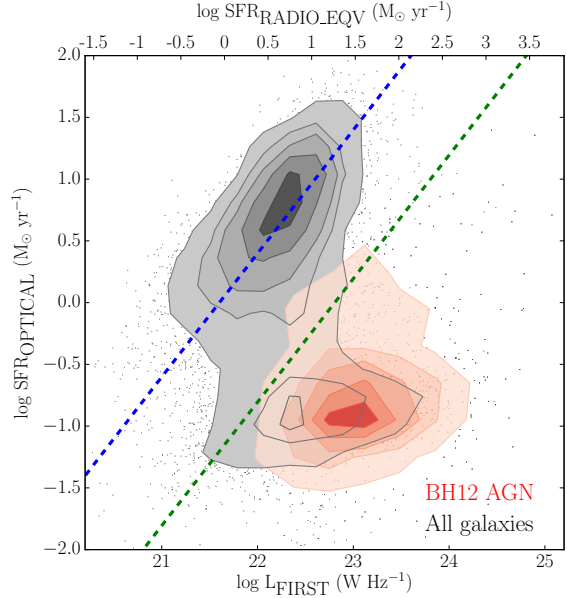

**Figure 1.** Comparison of the SFRs obtained via optical spectro-photometric data from the SDSS, with the FIRST 1.4 GHz luminosities (lower x-axis). Also shown is the 'radio-equivalent' SFRs (upper x-axis) calculated using the Condon (1992) relation from the FIRST luminosities. The blue dotted line shows where the radio luminosity is consistent with the MPA-JHU SFRs. The cloud of points clearly offset from this locus represents galaxies that exhibit radio excess i.e. the radio luminosity exceeds what is expected from star formation alone. The green dotted line separates the galaxies that have radio luminosities consistent with pure star formation from these radio-excess galaxies, which we classify as our faint AGN sample. The contours show regions containing 10, 30, 50, 70 and 90 per cent of the BH12 AGN (red) and all galaxies (black).

### 3.1 BPT classification

In Fig. 2, we show the same SFR comparison plot as in Fig. 1, but this time split by BPT classification. These classifications are obtained from B04, who used standard emission-line ratios, log([NII]/H$\alpha$) and log([OIII]/H$\beta$), measured from SDSS spectra, to assign each galaxy to one of six groups:

(i) 'star-forming': Galaxies with S/N > 3 in all four emission lines and which lie in the star-forming region of the BPT diagram.

(ii) 'low S/N star-forming': galaxies that have S/N > 2 in their SDSS emission lines but do not satisfy the criteria for AGN, composites or low S/N AGN.

(iii) 'composite': Galaxies that fall in the intermediate region of the BPT diagram, between purely star-forming systems and AGN, and have S/N > 3 in all four emission lines used in the BPT diagram. They are considered to have contributions from both an AGN and star formation activity.

(iv) 'AGN (non-LINER)': Galaxies that lie in the region above the theoretical upper limit for starburst models and hence must have a significant AGN contribution to their emission-line fluxes.

(v) 'low S/N LINER': Galaxies that lie above the theoretical limits for pure starburst models, but which have too low S/N in [OIII]$\lambda$5007 and/or H$\alpha$ emission lines to be included in the AGN classification.

(vi) 'unclassified': These are mainly sources that have very

MNRAS **000**, 1–10 (2017)*Faint local radio AGN* 3

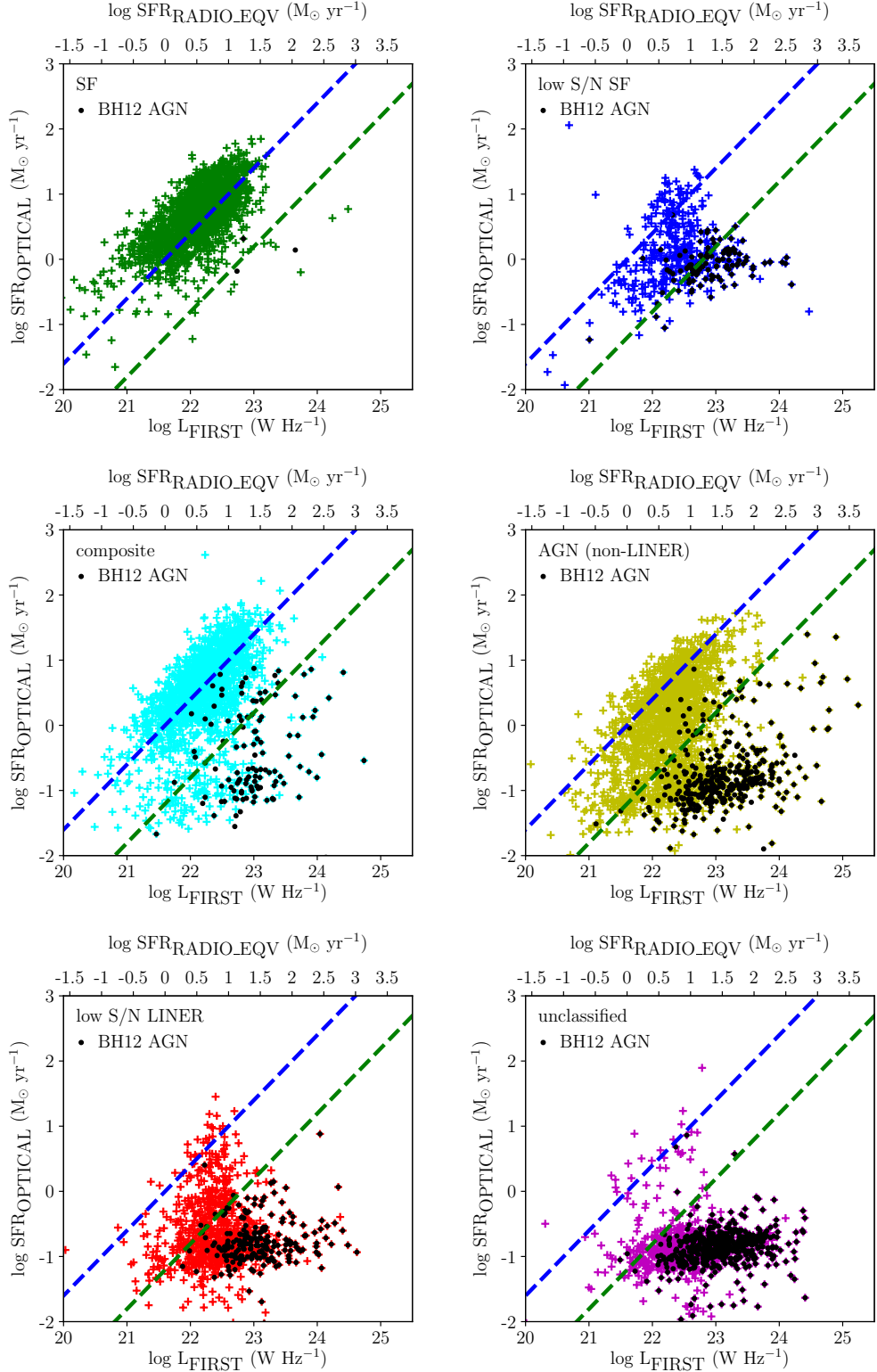

**Figure 2.** Same as Figure 1 but with each panel showing galaxies in specific BPT classes: star forming (top-left), low S/N star forming (top-right), composites (middle-left), AGN (middle-right), low S/N LINERs (bottom-left) and unclassified galaxies (bottom-right). In each panel, the BH12 AGN are shown using the black points. As in Figure 1, the blue dotted line is where the optical SFR and radio luminosity are consistent and the green dotted line indicates our AGN cut i.e. galaxies that fall to the right of this line are classified as our faint AGN sample.

**Table 1.** The first 20 objects in our faint radio AGN catalogue (the full table is available electronically). Columns are as follows: (1) RA (2) DEC (3) SDSS spectroscopic redshift (4) radio-equivalent SFR (5) optical SFR (6) stellar mass (7) bulge-to-total light ratios from Lackner & Gunn (2012) (8) dark matter halo mass from Yang et al. (2007) (9) BPT classifications from Brinchmann et al. (2004): Unclassified, SF = star-forming, low S/N SF = low S/N star-forming, Composite, AGN = AGN (Non-LINER), LINER.

| RA (deg) | Dec (deg) | z | log SFR$_{FIRST}$ (M$_\odot$yr$^{-1}$) | log SFR$_{MPA-JHU}$ (M$_\odot$yr$^{-1}$) | log M$_*$ (M$_\odot$) | B/T | log M$_{Halo}$ (M$_\odot$) | BPT |
|---|---|---|---|---|---|---|---|---|
| 0.3843786 | 14.185531 | 0.092 | 1.708 | 0.089 | 11.085 | | 12.359 | low S/N SF |
| 0.51508766 | 0.63190347 | 0.079 | 1.042 | -0.946 | 11.103 | | 12.423 | Unclassified |
| 0.7044455 | 0.7513351 | 0.087 | 1.371 | -0.228 | 11.249 | | 13.048 | AGN |
| 0.8844293 | 0.88283455 | 0.095 | 1.167 | -0.871 | 10.945 | | | LINER |
| 1.7523416 | 0.90292877 | 0.093 | 0.957 | -0.640 | 11.383 | | 13.084 | Unclassified |
| 2.2982507 | −0.6152013 | 0.073 | 2.450 | -0.036 | 11.267 | | 13.632 | AGN |
| 2.9384234 | −0.90852875 | 0.048 | 0.783 | -0.529 | 10.222 | | 11.571 | LINER |
| 3.8852134 | 14.659856 | 0.084 | 1.895 | -0.988 | 11.234 | | 13.604 | Unclassified |
| 4.1928964 | 1.2103353 | 0.044 | 0.391 | -1.179 | 11.051 | 1.0 | | AGN |
| 4.8955073 | 0.20058233 | 0.064 | 1.25 | -0.397 | 11.041 | | 12.378 | low S/N SF |
| 4.9122124 | −9.673885 | 0.084 | 1.411 | -0.404 | 11.690 | | 13.626 | AGN |
| 5.1146235 | 0.83355623 | 0.018 | -0.045 | -1.587 | 10.685 | | 11.883 | AGN |
| 5.1372666 | −0.14639343 | 0.064 | 1.047 | -0.715 | 11.423 | | 13.076 | AGN |
| 6.943303 | −10.262982 | 0.048 | 0.872 | -1.026 | 10.972 | 0.646 | 12.325 | Composite |
| 7.255976 | −1.2316802 | 0.082 | 1.268 | -2.009 | 10.285 | | 13.244 | Unclassified |
| 7.3684497 | −0.21260767 | 0.060 | 1.199 | -0.784 | 11.567 | | 14.208 | Unclassified |
| 7.4370885 | 0.16973962 | 0.060 | 1.292 | -0.116 | 10.227 | | 14.208 | AGN |
| 10.363505 | 15.049573 | 0.067 | 1.692 | -0.150 | 10.740 | | 12.115 | AGN |
| 10.686169 | −9.5545225 | 0.054 | 0.824 | -1.028 | 10.982 | | 14.530 | Unclassified |
| 10.792154 | -9.861478 | 0.050 | 0.630 | -0.899 | 11.276 | | 13.324 | AGN |

weak or no emission lines and hence could not be placed onto the BPT diagram.

Not unexpectedly, galaxies classified by the BPT analysis as star-forming fall overwhelmingly in the region where the radio luminosities are consistent with the optical SFRs (Fig. 2; top left), as would be expected from purely star-forming galaxies, with no (or a negligible) AGN contribution to the radio luminosity. Galaxies in this BPT class make up only 2 per cent of our faint AGN sample. Similarly, while a few galaxies in the low S/N star-forming class (Fig. 2; top right) have enhanced radio emission, the overwhelming majority do not and galaxies in this BPT class account for only 4 per cent of our faint AGN sample. The majority of the composite galaxies (Fig. 2, middle left), also lie in the region where the radio luminosity and optical SFRs agree. However, an increasing fraction of these galaxies show radio excess and, as a result, composites account for 11 per cent of the faint AGN. Not surprisingly, a much higher fraction of galaxies that fall in the AGN (non-LINER) class (Fig. 2; middle right) have a radio luminosity excess, with these galaxies accounting for 33 per cent of our faint AGN. Similarly, a majority of galaxies classified as low S/N LINER (Fig. 2; bottom left) show enhanced radio emission and this BPT class accounts for 25 per cent of our catalogue. Finally, almost all galaxies in the 'unclassified' BPT class (Fig. 2; bottom right) show a radio luminosity excess. While some (∼48 per cent) of these objects have been identified by BH12, many have not been previously identified as having an AGN component - in total there are 550 BPT-unclassified galaxies in our sample which are not in the BH12 catalogue. This BPT class accounts for 25 per cent of our faint AGN sample.

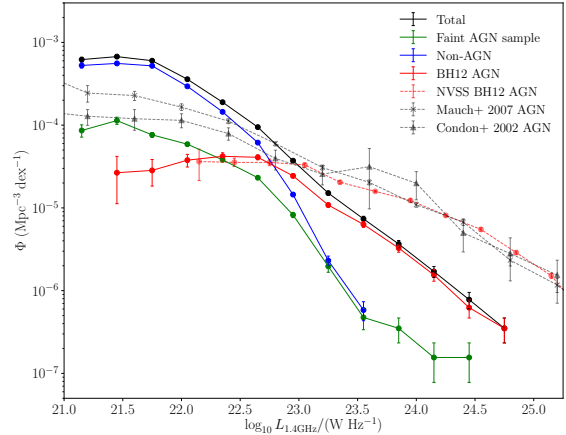

**Figure 3.** The local 1.4 GHz luminosity function, derived from FIRST fluxes. We present separate luminosity functions for the total 1.4 GHz radio luminosity (black), the faint radio AGN identified in this study (green), the AGN identified in BH12 (red) and non-AGN galaxies (blue). The luminosity functions are presented in Table 2. Also shown for comparison are the luminosity functions from NVSS data by BH12 (red dashed line), Mauch & Sadler (2007) (crosses) and Condon et al. (2002) (triangles). Our faint AGN make a non-negligible contribution to the local radio luminosity, particularly at low luminosities i.e. below $10^{22.5}$ W Hz$^{-1}$. The errors shown indicate Poisson errors.

## 4 RADIO LUMINOSITY FUNCTION

To investigate the contribution of the faint AGN to the total radio luminosity in the local Universe, we calculate the 1.4 GHz luminosity function (see Fig. 3). The luminosity functions are calculated using the standard Vmax method (Schmidt 1968; Felten

**Table 2.** The local ($z < 0.1$) radio luminosity functions at 1.4 GHz from the FIRST survey. The first column shows the ranges of 1.4 GHz luminosity included in each bin. The second, fourth and sixth and eight columns show the number of radio sources in each luminosity bin for the total, non-AGN, BH12 AGN and our faint AGN samples respectively. The third, fifth, seventh and ninth columns show the radio luminosity density, in Mpc$^{-3}$ dex$^{-1}$ for each of these samples. Note that the discrepancy between the total number of galaxies in star-forming, BH12 and faint AGN samples and the number in All radio sources results from the 37 galaxies that fall below our cut but were identified as AGN by BH12.

| log $L_{FIRST}$ $^a$ | All radio sources | | Star-forming | | BH12 AGN | | Faint AGN | |
|---|---|---|---|---|---|---|---|---|
| | N | $\Phi^a$ | N | $\Phi^a$ | N | $\Phi^a$ | N | $\Phi^a$ |
| 21-21.3 | 248 | $6.20 \times 10^{-4}$ | 211 | $5.27 \times 10^{-4}$ | 1 | - | 35 | $8.62 \times 10^{-5}$ |
| 21.3-21.6 | 680 | $6.73 \times 10^{-4}$ | 578 | $5.57 \times 10^{-4}$ | 3 | $2.66 \times 10^{-5}$ | 99 | $1.14 \times 10^{-4}$ |
| 21.6-21.9 | 1388 | $6.01 \times 10^{-4}$ | 1168 | $5.21 \times 10^{-4}$ | 8 | $2.84 \times 10^{-5}$ | 210 | $7.63 \times 10^{-5}$ |
| 21.9-22.2 | 2479 | $3.59 \times 10^{-4}$ | 2040 | $2.95 \times 10^{-4}$ | 33 | $3.78 \times 10^{-5}$ | 404 | $5.89 \times 10^{-5}$ |
| 22.2-22.5 | 3271 | $1.89 \times 10^{-4}$ | 2498 | $1.45 \times 10^{-4}$ | 93 | $4.20 \times 10^{-5}$ | 663 | $3.81 \times 10^{-5}$ |
| 22.5-22.8 | 2195 | $9.47 \times 10^{-5}$ | 1441 | $6.15 \times 10^{-5}$ | 230 | $4.09 \times 10^{-5}$ | 519 | $2.32 \times 10^{-5}$ |
| 22.8-23.1 | 885 | $3.72 \times 10^{-5}$ | 338 | $1.45 \times 10^{-5}$ | 349 | $2.44 \times 10^{-5}$ | 188 | $8.23 \times 10^{-6}$ |
| 23.1-23.4 | 374 | $1.51 \times 10^{-5}$ | 57 | $2.31 \times 10^{-6}$ | 274 | $1.08 \times 10^{-5}$ | 43 | $1.97 \times 10^{-6}$ |
| 23.4-23.7 | 185 | $7.37 \times 10^{-6}$ | 15 | $5.84 \times 10^{-7}$ | 158 | $6.31 \times 10^{-6}$ | 12 | $4.74 \times 10^{-7}$ |
| 23.7-24 | 94 | $3.66 \times 10^{-6}$ | 1 | - | 84 | $3.27 \times 10^{-6}$ | 9 | $3.50 \times 10^{-7}$ |
| 24-24.3 | 43 | $1.70 \times 10^{-6}$ | 0 | - | 39 | $1.55 \times 10^{-6}$ | 4 | $1.56 \times 10^{-7}$ |
| 24.3-24.6 | 20 | $7.78 \times 10^{-7}$ | 0 | - | 16 | $6.23 \times 10^{-7}$ | 4 | $1.56 \times 10^{-7}$ |
| 24.6-24.9 | 9 | $3.50 \times 10^{-7}$ | 0 | - | 9 | $3.50 \times 10^{-7}$ | 0 | - |

$^a$ in W Hz$^{-1}$
$^b$ Units of Mpc$^{-3}$ dex$^{-1}$

1977; Condon 1989), with

$$\rho = \Sigma_i \frac{1}{V_i} \qquad (4)$$

where $\rho$ is the number density of galaxies in each luminosity bin, and $V_i$ is the volume within which each source could be detected. This volume is calculated by finding the maximum and minimum redshifts at which an object with the luminosity of each source can be detected, given the redshift and magnitude limits used in the sample selection. These limits include our redshift criteria of $z < 0.1$, the magnitudes limits of the SDSS ($14.5 < r < 17.77$) and the FIRST survey source detection limit of 1 mJy. The area of the sky covered by both SDSS and FIRST is approximately 10,000 square degrees.

The total radio luminosity function for all radio sources at $z < 0.1$ is shown in Fig. 3 and Table 2, along with separate luminosity functions calculated for non-AGN, AGN identified by BH12 and for our faint AGN sample. The uncertainties quoted for these luminosity functions are statistical Poisson errors. Fig. 3 shows that the galaxies identified in this work contribute significantly to the total local radio luminosity, particularly at low radio luminosities i.e. below $10^{22.5}$ WHz$^{-1}$, where they contribute more than previously identified AGN in the BH12 sample. In particular, the faint AGN account for ~13 per cent of the radio luminosity budget at $z < 0.1$. This demonstrates that even if bright AGN have been removed from a sample, it can be unsafe to infer SFRs directly from radio emission *alone* as there is a significant population of faint radio AGN which may contaminate the radio SFRs.

We also show the AGN luminosity function calculated by BH12 using NVSS data. While the FIRST survey allows us to probe to fainter luminosities, it has the disadvantage of potentially resolving out extended emission. This is indicated by the difference between the luminosity functions for the BH12 sample calculated from FIRST in this work and the luminosity function presented in BH12 using NVSS. However, it is unlikely that we are missing significant flux in our FIRST analysis as sources in this luminosity range are compact as shown by the agreement of the NVSS and FIRST luminosity functions below $\sim 10^{23}$ WHz$^{-1}$.

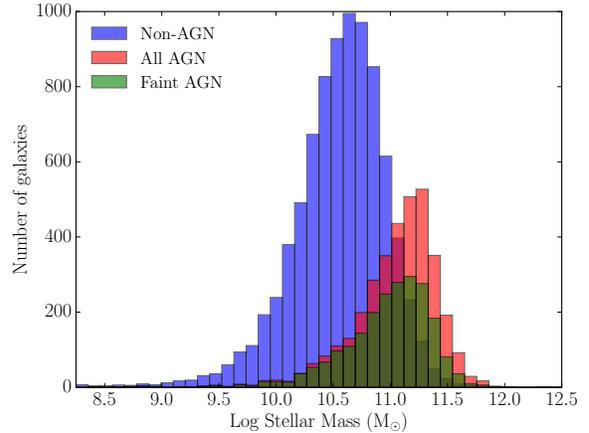

**Figure 4.** The distribution of stellar mass for non-AGN galaxies (blue), all AGN i.e. our faint AGN + the BH12 AGN (red) and the faint AGN sample only (green). Both the BH12 AGN and the faint AGN identified in this study peak at higher stellar masses than the non-AGN galaxies.

## 5 HOST GALAXY PROPERTIES

We complete our study by investigating key properties of the host galaxies of our faint AGN: stellar masses (Section 5.1), local environments (Section 5.2) and morphologies (Section 5.3).

## 5.1 Stellar Mass

Fig. 4 shows a histogram of the stellar masses of the faint AGN, all local AGN (defined as our faint AGN plus the BH12 AGN) and galaxies without AGN. The faint AGN have a median stellar mass of $10^{11.04\pm0.02} M_\odot$, comparable to that of the total AGN population, including BH12 AGN ($10^{11.11\pm0.02} M_\odot$). The errors on the medians have been calculated via a bootstrapping method. The faint AGN median mass is higher than the median stellar mass of the non-AGN galaxies $10^{10.59\pm0.02} M_\odot$. This is in general agreement with previous work which has found that AGN, and radio-loud AGN in particular, are typically hosted by more massive galaxies than their non-active counterparts (e.g. Kauffmann et al. 2003; Best et al. 2005; Heinis et al. 2016).

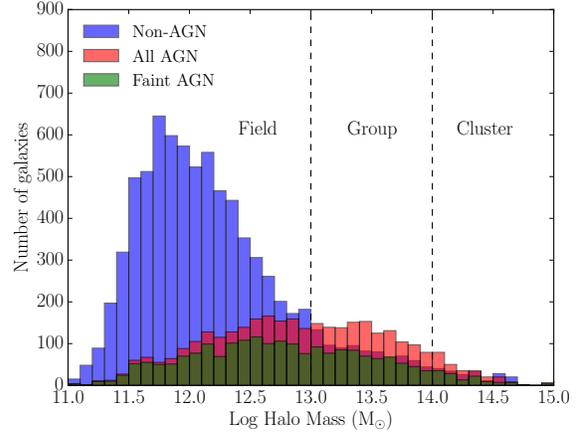

**Figure 5.** The distribution of the dark matter halo masses ($M_h$), which we use as a proxy for local environment, for the non-AGN (blue), all AGN i.e. our faint AGN + the BH12 AGN (red) and the faint AGN sample only (green). Galaxies with $M_h < 10^{13} M_\odot$ are considered to be in the field, those with halo masses in the range $10^{13} M_\odot < M_h < 10^{14} M_\odot$ are considered to be in groups, while cluster galaxies are defined as those with $M_h > 10^{14} M_\odot$. Both the BH12 AGN and the faint AGN identified in this work are typically found in higher density environments than the average non-AGN galaxy.

## 5.2 Environment

We next investigate the local environment of our faint AGN. Recall that cross-matching the Yang et al. (2007) catalogue to the SDSS-FIRST matched galaxies yields 11,313 objects with environment measurements, of which 3,300 are selected by our cut as AGN. Of these, 1,221 are included in the BH12 catalogue, resulting in 2,079 galaxies that are included in the faint AGN sample. Fig. 5 shows the distribution of halo masses for each of the samples as in Section 5.1: the faint AGN, all AGN and the non-AGN galaxies. We find that the non-AGN are mostly found in lower mass halos, i.e. in the field ($M_{halo} < 10^{13} M_\odot$), with a median halo mass of $10^{12.09\pm0.03} M_\odot$. This agrees with previous work, which finds that the vast majority of local star formation is in field galaxies (Lofthouse et al. 2017) and that there is a well-established relationship between the average SFR of a galaxy and its environment, such that higher SFRs are typically found in less dense environments (Hashimoto et al. 1998; Gómez et al. 2003; Kauffmann et al. 2004; Rasmussen et al. 2012; Wetzel et al. 2014). Galaxies in both the faint AGN sample and the total AGN sample are found over a much wider range of environments and peak at higher halo masses of $10^{12.75\pm0.04} M_\odot$ and $10^{12.90\pm0.05} M_\odot$ respectively.

It is worth noting that there is some debate in the literature about the dependence of the AGN fraction on environment. Many studies have suggested that local environment may not correlate strongly with the incidence of AGN, or have reported a decreasing fraction of AGN in denser environments (e.g. Miller et al. 2003; Ho et al. 2003; Kauffmann et al. 2004; Constantin & Vogeley 2006; Sabater et al. 2015). However, these studies have largely focused on *very* local environmental measures, such as the 2-point correlation function or nearest neighbour counts. Other studies which focus on the larger-scale environment, e.g. 'field', 'groups' and 'clusters' as is the case in this study, have found similar results to those shown here (Manzer & De Robertis 2014). For example, Best (2004) finds that radio-loud AGN are indeed preferentially hosted by galaxies in groups and poor-to-moderate richness galaxy clusters. Another factor which could explain such discrepancies in the literature is the method used to select the AGN. While some studies find no environmental dependence for optical, X-ray or mid-infrared selected AGN, radio-selected AGN, as in this study, have typically been reported to favour denser environments (Hickox et al. 2009; Argudo-Fernández et al. 2016), in agreement with our findings.

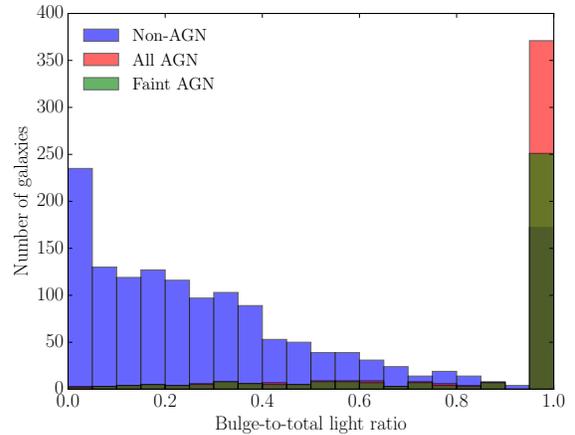

**Figure 6.** The distribution of bulge-to-total light ratios, which we use as a proxy for galaxy morphology, for the non-AGN galaxies (blue), all AGN i.e. our faint AGN + the BH12 AGN (red) and the faint AGN sample only (green). Non-AGN galaxies have predominantly low B/T values, i.e are morphological spirals, while almost all BH12 AGN and the faint identified in this work have B/T=1, i.e are morphological elliptical galaxies.

## 5.3 Morphology

The final host-galaxy property we investigate is morphology. We use the bulge-to-total light ratio, from Lackner & Gunn (2012), as a proxy for galaxy morphology. Recall that cross-matching this catalogue to the SDSS-FIRST matched galaxies yields 1971 sources with B/T measurements, of which 477 are AGN and 345 are in our sample. In Lofthouse et al. (2017), it was shown via visual inspection of a subsample of galaxies that these bulge to disk decompositions provide a reasonable method to split our galaxy sample into the typical elliptical and spiral morphologies. High B/T (B/T>0.5) values relate to morphological elliptical galaxies with B/T=1 indicating a pure elliptical with no disk component. Low B/T values (B/T<0.5) are morphological spirals, with B/T=0 indicating a bulgeless galaxy.

Fig. 6 shows the distribution of B/T ratios for each of the three samples. The non-AGN are found across the whole range of B/T ratios. They typically have relatively low values, with 76 per cent having B/T< 0.5, i.e they tend to be disk galaxies, in agreement with previous studies which show that the majority of star formation in the local Universe occurs in galaxies with late-type morphologies, (e.g. Kennicutt 1998; Kaviraj 2014a,b) and that galaxies classified as star-forming typically have disk-like morphologies (Brinchmann et al. 2004). The number of non-AGN decreases steadily as B/T increases, however there is a significant population (12 per cent of the non-AGN sample) with B/T= 1, i.e. pure elliptical galaxies. Of the 172 pure ellipticals that we classify as non-AGN, only 3 are unclassifiable via the BPT analysis. 106 are BPT-classified as composite, AGN(non-LINER) or low S/N LINER implying that there is some AGN component in these galaxies but the radio emission is not significantly enhanced to be selected by our method. The remaining 63 non-AGN pure ellipticals are classified by the BPT analysis as either star-forming or low S/N star-forming, which is consistent with these galaxies being truly non-AGN.

For the AGN samples, we find that there are few sources with disk-like morphologies (89 per cent have B/T $\geq$ 0.5) and the majority of the galaxies have B/T= 1, 78 per cent for the total AGN sample and 73 per cent for the faint AGN sample. This also agrees with previous studies that show that radio-loud AGN are preferentially found in massive elliptical galaxies (Wilson & Colbert 1995; Dunlop et al. 2003; Kauffmann et al. 2003, 2008).

We complete our analysis by briefly checking our morphological results with the local colour-magnitude relation. Using the k-corrected g-band and r-band magnitudes from the MPA-JHU catalogue, we split the sample into red and blue galaxies. We then recreate Fig. 1 showing the density of red sequence and blue cloud galaxies (Fig. 7). Galaxies identified as AGN (both the BH12 AGN and those in our faint AGN sample) are typically found on the red sequence which is dominated by early-type galaxies (e.g. Bower et al. 1992; Kodama & Arimoto 1997; Blanton & Moustakas 2009), in agreement with our results using the bulge-to-total light ratios which find that the majority of AGN have early-type morphology. In contrast, the non-AGN galaxies cover a much broader region of the colour-magnitude diagram and are predominantly in the blue cloud which is typically dominated by star-forming spiral galaxies (e.g. Tully et al. 1998; Baldry et al. 2004). This is again in agreement with our morphological results where we find that the non-AGN galaxies are disc-dominated.

Combined with our analysis of stellar mass and local environment above, we conclude that, on average, the faint AGN identified in this work are intermediate to high mass galaxies, reside across a range of environments and generally have early-type morphologies. While these AGN are typically lower luminosity than those previously identified by BH12, the properties of their host galaxies show the same trends across all of the variables studied here.

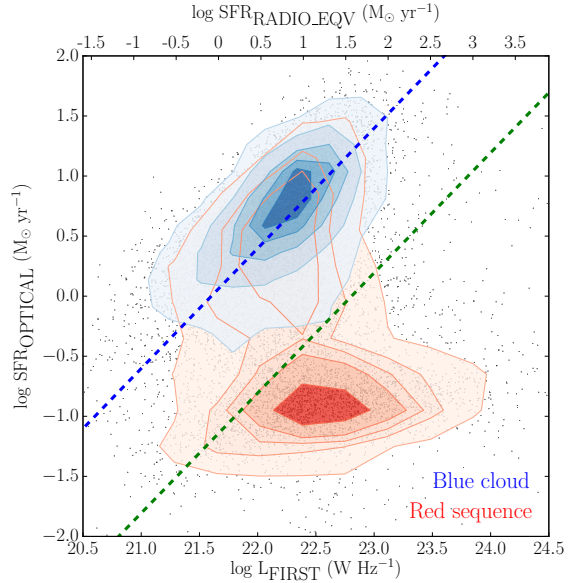

**Figure 7.** Comparison of the SFRs obtained via optical spectro-photometric data from the SDSS, with the FIRST 1.4 GHz luminosities as in Fig. 1 but coloured by position in the colour-magnitude diagram. Galaxies that are found in the blue cloud are mostly purely star-forming galaxies (shown in blue). However, a large fraction of red sequence galaxies (shown in red) have a radio luminosity excess and are classified as AGN. The contours show regions containing 10, 30, 50, 70 and 90 per cent of the galaxies in each colour.

## 6 SUMMARY

We have constructed a catalogue of local ($z < 0.1$) galaxies that host faint radio AGN. Our sample has been selected by identifying galaxies that exhibit an excess in their radio luminosities compared to what might be expected from the observed level of star formation activity in these systems. To do this, we have compared the measured 1.4 GHz luminosities to an optical SFR derived from SDSS spectro-photometry. In galaxies where the radio luminosity is driven by star formation, the radio luminosity is consistent with the optical SFRs. However, in galaxies where there is an AGN the radio luminosity will exhibit an excess relative to that expected based on the optical SFR, enabling us to identify systems that host an AGN. While some of these AGN (those that have relatively high radio luminosities) have been identified by BH12, a majority have not been previously identified. Our results demonstrate that it is unsafe to infer SFRs from radio emission *alone* even if bright AGN have been excluded from a sample since there is a significant population of faint radio AGN which may contaminate the radio SFRs.

We refer to the new AGN identified in this study as faint AGN. Our main results can be summarized as follows:

• Galaxies with radio excess are typically classified as AGN, LINERs or composite galaxies from their optical emission line ratios. However, $\sim$25 per cent of the catalogue are 'unclassified' due to the absence or low S/N of these emission lines.

• A calculation of the local radio luminosity functions separately for the faint AGN, all AGN (our faint AGN + BH12 AGN) and non-AGN galaxies indicates that the faint AGN provide a significant contribution to the local 1.4 GHz luminosity function, particularly at low radio luminosities and contribute 13 per cent of the local radio luminosity budget.

- The host galaxies of our faint AGN are predominantly high stellar mass galaxies, with a median stellar mass of $10^{11.04\pm0.02}M_\odot$. This is higher than the median stellar mass for non-AGN systems ($10^{10.59\pm0.02}M_\odot$).
- The faint AGN span a wide range of environments, from the field to clusters, but are typically in denser environments than non-AGN galaxies.
- ∼86 per cent of the faint AGN galaxies are bulge-dominated systems with ∼73 per cent pure elliptical galaxies.

This study demonstrates a general technique by which radio-loud AGN can be identified in galaxy populations where reliable optical SFRs can be extracted using spectro-photometry and where radio data are available so that a radio excess can be measured. The radio luminosity and optical SFR can then be compared to identify galaxies which exhibit a radio excess. For example, deep-wide radio surveys, such as those using SKA precursors, can be combined with existing optical datasets like the SDSS or forthcoming surveys like LSST to extract catalogues of AGN, via such a technique, across a large range in redshift.

**ACKNOWLEDGEMENTS**

We are grateful to the anonymous referee for many constructive comments that improved the original version of this paper. EKL acknowledges support from the UK's Science and Technology Facilities Council [grant number St/K502029/1]. SK acknowledges a Senior Research Fellowship from Worcester College Oxford. MJH is supported by STFC grant ST/M001008/1.

This paper has been typeset from a TeX/LaTeX file prepared by the author.